\documentclass[twocolumn,
 showpacs,  
 preprintnumbers,  
 aps,  
 prd,  
 a4paper,  
 nofootinbib,  
 tightenlines,  
 floats,floatfix  
 ]{revtex4}

\usepackage{graphicx, epsfig}
\usepackage{color}
\usepackage{mathrsfs}

\usepackage{amsmath}
\usepackage{amssymb}

\numberwithin{equation}{section}
\usepackage{epsfig}

\def \lleq {\lower0.9ex\hbox{ $\buildrel < \over \sim$} ~}
\def \ggeq {\lower0.9ex\hbox{ $\buildrel > \over \sim$} ~}

\newcommand{\ben}{\begin{eqnarray}}
\newcommand{\een}{\end{eqnarray}}

\def \omms   {\Omega_m}

\def \omm  {\Omega_{0 {\rm m}}}

\def \beq  {\begin{equation}}
\def \eeq  {\end{equation}}
\def \ber  {\begin{eqnarray}}
\def \eer  {\end{eqnarray}}

\newcommand {\ga} {\ {\raise-.5ex\hbox{$\buildrel>\over\sim$}}\ }
\newcommand {\la} {\ {\raise-.5ex\hbox{$\buildrel<\over\sim$}}\ }

\renewcommand{\(}{\left(}
\renewcommand{\)}{\right)}
\renewcommand{\[}{\left[}
\renewcommand{\]}{\right]}

\begin{document}
\newcommand{\newc}{\newcommand}

\newc{\be}{\begin{equation}}
\newc{\ee}{\end{equation}}
\newc{\ba}{\begin{eqnarray}}
\newc{\ea}{\end{eqnarray}}
\newc{\bea}{\begin{eqnarray*}}
\newc{\eea}{\end{eqnarray*}}
\newc{\D}{\partial}
\newc{\ie}{{\it i.e.} }
\newc{\eg}{{\it e.g.} }
\newc{\etc}{{\it etc.} }
\newc{\etal}{{\it et al.}}
\newc{\lcdm }{$\Lambda$CDM }
\newcommand{\nn}{\nonumber}
\newc{\ra}{\rightarrow}
\newc{\lra}{\leftrightarrow}
\newc{\lsim}{\buildrel{<}\over{\sim}}
\newc{\gsim}{\buildrel{>}\over{\sim}}

\title{Evolution of Dark Energy Perturbations in Scalar-Tensor Cosmologies}
\author{J. C. Bueno Sanchez and L. Perivolaropoulos}
\affiliation{Department of Physics, University of Ioannina, Greece}
\date{\today}
\begin{abstract}
We solve analytically and numerically the generalized Einstein equations in scalar-tensor cosmologies to obtain the evolution of dark energy and matter linear perturbations. We compare our results with the corresponding results for minimally coupled quintessence perturbations. Our results for natural ($O(1)$) values of parameters in the Lagrangian which lead to a background expansion  similar to \lcdm are summarized as follows:  1. Scalar-Tensor dark energy density perturbations are amplified by a factor of about $10^4$ compared to minimally coupled quintessence perturbations on scales less than about $1000{\rm h^{-1} Mpc}$ (sub-Hubble scales). This amplification factor becomes even larger ($\gtrsim 10^6$) for scales less than $100{\rm h^{-1} Mpc}$. On these scales dark energy perturbations constitute a fraction of about $10\%$ compared to matter density perturbations. 2. Scalar-Tensor dark energy density perturbations are anti-correlated with matter linear perturbations on sub-Hubble scales. Thus clusters of galaxies are predicted to overlap with voids of dark energy. 3. This anti-correlation of matter with negative pressure perturbations induces a mild amplification of matter perturbations by about $10\%$ on sub-Hubble scales. 4. The evolution of scalar field perturbations on sub-Hubble scales, is scale independent and therefore it corresponds to a vanishing effective speed of sound ($c_{s\Phi}=0$). It also involves large oscillations at early times induced by the amplified effective mass of the field. This mass amplification is due to the non-minimal coupling of the field to the Ricci curvature scalar and (therefore) to matter. No such oscillations are present in minimally coupled quintessence perturbations which are suppressed on sub-Hubble scales ($c_{s\Phi}=1$). We briefly discuss the observational implications of our results which may include predictions for galaxy and cluster halo profiles which are modified compared to \lcdm. The observed properties of these profiles are known to be in some tension with the predictions of \lcdm.
\end{abstract}
\pacs{98.80.Es,98.65.Dx,98.62.Sb}
\maketitle

\section{Introduction}
A wide range of cosmological observations indicate that the universe has entered a phase of accelerating expansion. These observations include both direct geometric probes of the expanding FRW metric and dynamical probes of the growth rate of matter perturbations. This growth depends on both the expansion rate and the gravitational law on large scales.

Geometric probes of the cosmic expansion include the following:
\begin{itemize}
\item Type Ia supernovae (SnIa) standard candles \cite{Hicken:2009dk,Kowalski:2008ez}
\item The angular location of the first peak in the CMB perturbations angular power spectrum\cite{Komatsu:2008hk}. This peak probes the integrated cosmic expansion rate using the last scattering horizon as a standard ruler.
\item Baryon acoustic oscillations of the matter density power spectrum. These oscillations also probe the integrated cosmic expansion rate on more recent redshifts using the last scattering horizon as a standard ruler \cite{Percival:2009xn}.
\item Other less accurate standard candles (Gamma Ray Bursts \cite{Basilakos:2008tp}, HII starburst galaxies\cite{Plionis:2009up}) and standard rulers (cluster gas mass fraction \cite{Allen:2004cd}) as well as probes of the age of the universe \cite{Krauss:2003em}.
\end{itemize}
Dynamical probes of the cosmic expansion and the gravitational law on cosmological scales include:
\begin{itemize}
\item X-Ray cluster growth data \cite{Rapetti:2009ri}
\item Power spectrum of Ly-$\alpha$ forest at various redshift slices \cite{McDonald:2004xn,Nesseris:2007pa}
\item Redshift distortion observed through the anisotropic pattern of galactic redshifts on cluster scales \cite{Hawkins:2002sg,Nesseris:2007pa}
\item Weak lensing surveys \cite{Fu:2007qq,Benjamin:2007ys,Amendola:2007rr}
\end{itemize}
These cosmological observations converge on the fact that the simplest model describing well the cosmic expansion rate is the one corresponding to a cosmological constant \cite{lcdmrev} in a flat space namely:
\be H(z)^2=H_0^2 \[\omm (1+z)^3 + (1-\omm)\] \label{hzlcdm} \ee where $H(z)$ is the Hubble expansion rate at redshift $z$, $H_0=H(z=0)$ and $\omm$ the present matter density normalized to the present critical density for flatness. This form of $H(z)$ and other similar, more complicated forms of it that are also consistent with cosmological data, are predicted by three broad classes of models:
\begin{itemize}
\item {\bf Dark Energy Models} attribute the observed accelerating expansion to an unknown energy component (dark energy\cite{Copeland:2006wr,Frieman:2008sn,Perivolaropoulos:2006ce}) with negative pressure and repulsive gravitational properties which dominates the universe at recent cosmological times. The simplest model of this class is based on dark energy with constant energy density (in time and in space) {\it the cosmological constant\cite{lcdmrev}} \footnote{Vacuum energy of quantum fields could in principle provide such a constant energy density\cite{Perivolaropoulos:2008pg} albeit with a much larger magnitude than observed.}. This simple and physically motivated model however, is plagued by fine tuning problems \cite{Zlatev:1998tr}. Alternatively, dark energy may be described by a minimally coupled scalar field (quintessence \cite{Zlatev:1998tr} and k-essence\cite{ArmendarizPicon:2000ah}) which acquires negative pressure during an evolution dominated by potential energy. Quintessence models also require the introduction of a fine tuned unnaturally small mass scale comparable to the Hubble scale today. The role of dark energy can also be played by the introduction of perfect fluids with proper equation of state parameters \cite{Carturan:2002si} (eg Chaplygin gas \cite{Bilic:2001cg,Bento:2002ps}) but this approach is less motivated physically.
\item {\bf Modified Gravity Models} are based on large scale modifications of General Relativity (GR) which lead to repulsive gravitational properties and accelerate the cosmic expansion. Representatives of this class of models include $f(R)$ gravity \cite{Nojiri:2006ri,Amendola:2006we}, DGP models \cite{Dvali:2000hr} and Scalar-Tensor (ST) theories \cite{st-action,EspositoFarese:2000ij} (for a historical review of ST theories see \cite{Brans:2005ra}). These models also require some degree of fine tuning of parameters but they are more appealing from a physical point of view.
\item{\bf Local Void Models} assume the existence of an unusually large underdensity (void) \cite{Alnes:2005rw,Alexander:2007xx} on scales of about 1 Gpc which induces an apparent isotropic accelerating expansion to observers located close to the center of it  (within about $20\;{\rm Mpc}$). These models are in some sense minimal since they require no new theoretical input but they have two sources of fine tuning: the location of the observer at the center of the void and the statistically improbable assumption of the formation of a 1 Gpc void.
\end{itemize}
Scalar-Tensor (ST) cosmological models \cite{EspositoFarese:2000ij} (extended quintessence \cite{Perrotta:1999am}) constitute a fairly generic representative of modified gravity models. They are based on the promotion of Newton's constant to a non-minimally coupled to curvature scalar field whose dynamics is determined by a potential $U(\Phi)$ and by the functional form of the non-minimal coupling $F(\Phi)$.

The deviation of these models from GR is tightly constrained locally by solar system observations and by small scale gravitational experiments \cite{solsyst,EspositoFarese:2004cc}. These constraints however are significantly less stringent on cosmological scales \cite{Uzan:2009ri,Umezu:2005ee,Damour:1998ae} and may also be evaded by chameleon type arguments \cite{Khoury:2003rn}. These arguments are based on the fact that the effective mass of the non-minimally coupled scalar field depends on the local matter density which is naturally high on Earth and in the solar system where most of the constraints are obtained. Such a high mass freezes the dynamics of the scalar field locally and allows it to mimic GR on solar system scales. On cosmological scales however, where the mean density is low, the field is allowed to dynamically evolve and drive the observed accelerated expansion.

Another potential mechanism to relax the strong local constraints on ST theories involves the fact that GR is an attractor\cite{Damour:1993id} in the phase space of ST theories. An early cosmological deviation from GR is therefore allowed since it naturally leads to an evolution towards a late phase region that emulates GR in agreement with local space-time observations.

The main advantages of ST cosmologies may be summarized as follows:
\begin{itemize}
\item ST theories are natural alternatives to GR. Scalar partners to the graviton naturally arise in most attempts to quantize gravity or unify it with other interactions (eg the dilaton in superstring theories \cite{Barreiro:1998aj} or the radion in extra dimensional Kaluza-Klein\cite{Perivolaropoulos:2002pn} and brane cosmologies\cite{Csaki:1999mp}).
\item  Even though ST theories respect most of the symmetries of GR (local Lorentz invariance, conservation laws, weak equivalence principle) they are fairly general and several modified gravity theories may be viewed as effective special cases of ST theories (eg $f(R)$ gravity \cite{Chiba:2003ir}, Kaluza-Klein models \cite{Perivolaropoulos:2002pn} etc).
\item The direct coupling of scalar field to curvature (and thus to matter density) provides in principle a mechanism to evade the coincidence problem ie address the question `Why did the accelerating expansion start soon after matter domination in the universe \cite{Chimento:2003iea}'.
\item ST theories in contrast to minimally coupled quintessence, naturally allow superaccelerating expansion rate \cite{Boisseau:2000pr,Perivolaropoulos:2005yv}. Superacceleration corresponds to an effective dark energy equation of state $w_{\rm eff}<-1$. Thus, crossing of the phantom divide line $w=-1$ is naturally allowed in ST theories. Superacceleration is equivalent to violation of the inequality \be \frac{dH(z)^2}{dz}\geq 3\omm H_0^2 (1+z)^2 \label{pdlineq} \ee In contrast, this inequality is enforced in most models of GR with dark energy\cite{Vikman:2004dc} and corresponds to the null energy condition $\rho_{\rm eff} \geq -p_{\rm eff}$ where $\rho_{\rm eff}$, and $p_{\rm eff}$ are the effective energy and pressure driving the accelerated expansion \cite{Nesseris:2006er}.
\item The growth of matter and scalar field perturbations is distinct in ST cosmologies compared to those based on GR. In particular, on sub-Hubble scales the growth of matter perturbations is driven by an effective evolving gravitational constant \cite{Boisseau:2000pr,EspositoFarese:2000ij}. The perturbations of the non-minimally coupled scalar field are scale independent and not negligible on small scales \cite{EspositoFarese:2000ij,Perrotta:2002sw}. This is in contrast with the case of minimally coupled quintessence where field perturbations scale as $\delta \Phi \sim \frac{H(a)^2 a^2}{k^2}$ and are negligible on small sub-Hubble scales \cite{Weller:2003hw,Unnikrishnan:2008qe} ($\frac{k}{a} \gg H$). These amplified perturbations of non-minimally coupled ST dark energy (extended quintessence) trigger a local amplification in matter perturbations and could potentially provide a resolution to a puzzle of \lcdm cosmology related to the concentration parameter of cluster and galaxy dark matter halo profiles \cite{Gentile:2007sb}. These profiles are observed to be significantly more concentrated than predicted by \lcdm \cite{Broadhurst:2004bi,Umetsu:2007pq}. Amplified perturbations of dark energy could naturally lead to a resolution of this puzzle \cite{Basilakos:2009mz}.
\end{itemize}
The growth of perturbations in ST theories and its quantitative comparison with the corresponding growth in GR is the main focus of this study. This comparison can lead to the derivation of potential signatures of ST theories on the power spectrum and on other other observables related to the growth of density perturbations.

Linear dark energy perturbations have been extensively studied especially in the context of GR \cite{Weller:2003hw,Unnikrishnan:2008qe} and constrained by the CMB angular power spectrum spectrum\cite{Bean:2003fb,dePutter:2010vy}. In the context of minimally coupled quintessence it was found that the scalar field density perturbations exist on all scales but they are strongly scale dependent and negligible on sub-Hubble scales \cite{Unnikrishnan:2008qe}. An interesting anti-correlation between dark matter and quintessence perturbations has also been pointed out\cite{Dutta:2006pn}.  On larger (super-Hubble) scales, they can range up to about $10\%$ compared to the matter density perturbations and they leave a trace on the low $l$ multipoles of the CMB angular power spectrum through the ISW effect \cite{Weller:2003hw}. The suppression of sub-Hubble scalar field density fluctuations in minimally coupled quintessence originates from the value of the non-adiabatic speed of sound in these models which is about equal to unity \cite{Ferreira:1997hj,dePutter:2010vy}. The sound speed determines the sound horizon of the dark energy fluid, $l_s = c_s/H$. On scales below this sound horizon, the perturbations vanish while on scales above $l_s$ dark energy can cluster.

In ST theories, the non-minimal coupling of the scalar field to curvature perturbations (which in turn are driven by matter perturbations) leads to an amplification of the scalar field perturbations on sub-Hubble scales.  Thus, it may be shown that on sub-Hubble scales the field perturbations $\delta \Phi$ are scale independent \cite{EspositoFarese:2000ij} and therefore the effective speed of sound $c_{s\Phi}$ for ST field perturbations vanishes. As discussed below, the corresponding scalar field density perturbations are also amplified but they are {\it anti-correlated} with respect to matter perturbations ($\frac{\delta_\Phi(k,t_0)}{\delta_m(k,t_0)}<0$). In addition, as shown in the following sections the ratio of the scalar field density perturbations over the matter density perturbations on sub-Hubble scales is also independent of the scale and can become significant (up to about $10\%$) for cosmologically viable models. The goal of the present study is to demonstrate the above points using both qualitative analytical approximations and detailed numerical analysis.

The structure of this paper is the following: In section II we derive the main equations that determine the evolution of the background and the linear perturbations in ST cosmologies. We also use analytical approximations to derive a few qualitative features of the solutions for the field and matter perturbations in these theories and compare them with the corresponding features of GR solutions. In section III we present a detailed numerical solution of the linear perturbation equations using a specific form of the ST potentials which is able to produce an observationally viable expansion background similar to \lcdm. Finally, in section IV we summarize our main results and discuss future prospects and extensions of this study.

\section{Perturbations in Scalar-Tensor Cosmologies}
We consider the following ST action in the physical Jordan frame \cite{st-action,EspositoFarese:2000ij}
\begin{widetext}
\be
S={1\over 16\pi G} \int d^4x \sqrt{-g}
\Bigl(F(\Phi)~R -
Z(\Phi)~g^{\mu\nu}
\partial_{\mu}\Phi
\partial_{\nu}\Phi
- 2U(\Phi) \Bigr)
+ S_m[\psi_m; g_{\mu\nu}]\ .
\label{st-action}
\ee
\end{widetext}
where G is the bare gravitational constant, $R$ is the scalar curvature of the metric $g_{\mu\nu}$ and $S_m$ is the action of matter fields. In what follows we use units such that $8\pi G=1$. The variation of the dimensionless function $F(\Phi)$ describes the variation of the effective gravitational constant. This variation (spatial or temporal) is severely constrained by solar system experiments \cite{EspositoFarese:2004cc,solsyst}. The GR limit of ST theories is obtained either by fixing $F(\Phi)=\Phi_0\simeq 1$ ($\Phi_0$ is a constant) or by freezing the dynamics of $\Phi$ using the function $Z(\Phi)$ or the potential $U(\Phi)$. For example a large and steep $Z(\Phi)$ makes it very costly energetically for $\Phi$ to develop a kinetic term while a steep confining $U(\Phi)$ (massive $\Phi$) can make it very costly energetically for $\Phi$ to develop potential energy. In both cases we have an effective {\it freezing} of the dynamics which reduces the ST theory to GR.

Considering variation of the action (\ref{st-action}) we obtain the dynamical equations
\begin{widetext}
\ba
F(\Phi) \left(R_{\mu\nu}-{1\over2}g_{\mu\nu}R\right)
&=&  T_{\mu\nu}
+ Z(\Phi) \left(\partial_\mu\Phi\partial_\nu\Phi
- {1\over 2}g_{\mu\nu}
(\partial_\alpha\Phi)^2\right)
+\nabla_\mu\partial_\nu F(\Phi) - g_{\mu\nu}\Box F(\Phi)
- g_{\mu\nu} U(\Phi)\ ,
\label{stmetrevol}\\
2Z(\Phi)~\Box\Phi &=&
-{dF\over d\Phi}\,R - {dZ\over d\Phi}\,(\partial_\alpha\Phi)^2
+ 2 {dU\over d\Phi}\ ,
\label{stfevol}\\
\nabla_\mu T^\mu_\nu &=& 0\ ,
\label{tmncons}
\ea
\end{widetext}
where $T^{\mu \nu}$ is the
matter energy-momentum tensor $T^{\mu\nu} \equiv (2/\sqrt{-g})\times
\delta S_m/\delta g_{\mu\nu}$.

Considering a flat cosmological FRW background
\be
ds^2 = -dt^2 + a^2(t)\[ {dr^2}
+ r^2 \left(d\theta^2 + \sin^2\theta~d\phi^2\right)\]
\label{frw}
\ee
where matter is described by a pressureless perfect fluid with density and pressure $(\rho,p)=(\rho_m,0)$ we obtain the equations for the evolution of the background
\begin{eqnarray}
&&3F \cdot H^2
=
 \rho_m
+{1\over 2} \dot\Phi^2 - 3 H \dot F + U \equiv \rho_{tot}\
\label{stbh1}\\
&&-2F\cdot \dot H
=
\rho_m + \dot\Phi^2 +\ddot F - H\dot F \equiv \rho_{tot}+p_{tot}\
\label{stbh2}\\
&&\ddot\Phi+3H\dot\Phi
=
3{dF\over d\Phi}\left(\dot H + 2 H^2 \right)
- {dU\over d\Phi}\
\label{stbphi}\\
&&\dot\rho_m + 3H~\rho_m = 0
\label{stbmat}
\end{eqnarray}
where we have rescaled $\Phi$ so that $Z=1$ assuming that $Z>0$. As discussed in detail in the next section it is straightforward to solve this system numerically with initial conditions corresponding to the time of recombination ($z\simeq 1000$) and find the evolution of the background homogeneous field $\Phi(t)$, scale factor $a(t)$ and matter density $\rho_m(t)\sim a(t)^{-3}$. Notice that this is a system of {\it three} independent equations. In order to obtain them we substitute $\dot H$ from Eq.~ (\ref{stbh2}) in (\ref{stbphi}) and use Eqs.~(\ref{stbh1}), (\ref{stbphi}) and (\ref{stbmat}) for the numerical analysis of the background evolution. This substitution in (\ref{stbphi}) reveals a new effective mass for the field $\Phi$ which depends on the background matter density $\rho_m(t)$. More details on the rescaling and the initial conditions are provided in the next section.

 In order to obtain the evolution of perturbations we consider the perturbed FRW metric which in the Newtonian gauge takes the form
\be
d s^2 = - (1+2\phi)dt^2 +a^2 (1-2\psi)\delta_{ij}   dx^i dx^j \label{pertmet}
\ee
The linear gravitational potentials $\phi$ and $\psi$ along with the scalar field perturbations $\delta \Phi$ are sufficient to fully determine the cosmological perturbations in these theories. Using the perturbed metric (\ref{pertmet}) in the generalized Einstein Eqs.~(\ref{stmetrevol}) it is straightforward to find the connection between the potentials $\phi$ and $\psi$
\be
\phi=\psi - \frac{F_{,\Phi}}{F}\delta\Phi \label{psiphicon} \ee where $F_{,\Phi}\equiv \frac{dF}{d\Phi}$. For $F=1$ (GR case)  Eq.~ (\ref{psiphicon}) reduces to the well known relation $\phi=\psi$ which is valid in GR in the absence of anisotropic stresses. It is straightforward to obtain two additional differential equations for the perturbations $\phi$, $\psi$ and $\delta \Phi$ using Eq.~ (\ref{pertmet}) and perturbing $\Phi$ in (\ref{stmetrevol})-(\ref{stfevol}). These equations are of the form \cite{Hwang:2005hb,EspositoFarese:2000ij,Copeland:2006wr}
\begin{widetext}
\be {\dot \xi}+2H \xi + \left(3{\dot H}-\frac{k^2}{a^2}\right)\phi -\frac{1}{2}(\delta \mu + 3 \delta q)=0 \label{perts1} \ee
\be {\ddot{\delta \Phi}} + 3H{\dot{\delta \Phi}} + \left(\frac{k^2}{a^2} + U_{,\Phi\Phi}-\frac{R}{2} F_{,\Phi\Phi}\right)\delta\Phi- {\dot \Phi} {\dot \phi} - (2 {\ddot \Phi} + 3H {\dot \Phi})\phi - {\dot \Phi} \xi - \frac{F_{,\Phi}}{2}\delta R =0 \label{perts2} \ee
\end{widetext}
where \be \xi\equiv 3(H \phi + {\dot \psi}) \label{xidef} \ee
\be \delta \mu \equiv \frac{\delta \rho_{tot}}{F}= -2\left(H \xi + \frac{k^2}{a^2}\psi\right) \label{poisson} \ee corresponds to the generalized Poisson equation, \be R=6({\dot H} + 2H^2) \label{ricci} \ee is the Ricci curvature scalar, \be \delta R = 2\[-{\dot \xi} - 4 H \xi - \left(3 {\dot H}-\frac{k^2}{a^2}\right) \phi - 2 \frac{k^2}{a^2} \psi\]  \label{dricci} \ee is the Ricci scalar perturbation and \begin{widetext} \ba \delta q &\equiv&\frac{\delta p_{tot}}{F}= \frac{1}{F} \left[{\dot \Phi} {\dot {\delta \Phi}}+ \frac{1}{2}(F_{,\Phi}R-2U_{,\Phi})\delta \Phi +\frac{d^2}{dt^2}(F_{,\Phi}\delta\Phi)+2H\frac{d}{dt}(F_{,\Phi}\delta\Phi)+\left(-3H^2 -{\dot H}+\frac{2}{3}\frac{k^2}{a^2}\right)F_{,\Phi}\delta\Phi-\right. \nn \\ &-& \left.{\dot F}{\dot \phi}-{\dot \Phi}^2 \phi - 2{\ddot F} \phi - 2 H {\dot F} \phi - \frac{2}{3} {\dot F} \xi\right]  \label{dpdef} \ea
\end{widetext}
is the effective pressure perturbation. Assuming specific forms for the field potentials F and U in Eqs.~(\ref{perts1})-(\ref{perts2}), eliminating the gravitational potential $\phi$ using Eq.~ (\ref{psiphicon}) and using the background solution of the system (\ref{stbh1})-(\ref{stbmat}), we may numerically solve for the perturbations $\psi(k,t)$, $\delta\Phi(k,t)$. After proper rescaling we solve the system in the next section with initial conditions at recombination ($z\simeq 1000$) corresponding to an initially smooth scalar field $\Phi$ and a properly rescaled gravitational potential $\psi$ in a background with small initial deviation from GR. Assuming that $\psi(k,t)$ and $\delta\Phi(k,t)$ have been obtained by the numerical solution of the system (\ref{perts1})-(\ref{perts2}) in the background solution of (\ref{stbh1}), (\ref{stbphi}), (\ref{stbmat}), the gravitational potential $\phi(k,t)$ can be obtained by using Eq.~ (\ref{psiphicon}) that connects the two gravitational potentials in the context of ST theories.

It is straightforward to obtain the density perturbations of both dark energy ($\delta_\Phi\equiv \frac{\delta \rho_\Phi}{\rho_{tot}}$) and matter ($\delta_m\equiv \frac{\delta \rho_m}{\rho_{tot}}$) in terms of the numerically derived perturbations $\phi(k,t)$, $\psi(k,t)$ and $\delta\Phi(k,t)$ and the corresponding background. The total effective density perturbation $\delta \mu$ of Eq.~ (\ref{poisson}) may be expressed in terms of field and matter perturbations as \cite{Hwang:2005hb,Copeland:2006wr}
\begin{widetext}
\ba
&&\delta \mu\equiv \frac{\delta \rho_{tot}}{F}=\frac{\delta \rho_\Phi +\delta \rho_m}{F}=\nn \\
&&=\frac{1}{F}\left[{\dot \Phi}{\dot {\delta \Phi}}-\frac{F_{,\Phi} R \delta \Phi}{2} + U_{,\Phi} \delta \Phi - 3H \frac{d}{dt}(F_{,\Phi}\delta \Phi)+\left(3{\dot H}+3H^2-\frac{k^2}{a^2}\right)F_{,\Phi}\delta \Phi+(3H{\dot F}-{\dot \Phi}^2)\phi + \right.\nn \\&& \left.+{\dot F}\xi +\delta \rho_m-\frac{F_{,\Phi}}{F} \delta \Phi \rho_m\right] \label{dmu2} \ea
\end{widetext}
while the corresponding effective total density $\mu$ is expressed as (see also Eq.~ (\ref{stbh1}))
\be
\mu \equiv \frac{\rho_{tot}}{F}=\frac{\rho_\Phi+\rho_m}{F}=
  \frac{1}{F}\( \frac{1}{2}{\dot \Phi}^2+U-3H{\dot F} +\rho_m\) \label{mu} \ee

Using now the above discussed numerical solution for  the perturbations $\phi(k,t)$, $\psi(k,t)$ and $\delta\Phi(k,t)$ and Eqs.~(\ref{poisson}), (\ref{dmu2}), (\ref{mu}) we obtain the following density perturbations
\be
\delta \rho_{\Phi} = F \cdot \delta \mu \vert_{\delta \rho_m=0} \label{drhophi} \ee
from Eq.~ (\ref{dmu2}), \be \rho_{tot} = F \cdot \mu \label{rhotot} \ee from Eq.~ (\ref{mu}), \be \delta_\Phi=\frac{\delta \rho_\Phi}{\rho_{tot}}= \frac{ \delta \mu \vert_{\delta \rho_m=0}}{\mu} \label{phiperts} \ee from Eqs.~(\ref{drhophi}), (\ref{rhotot}), \be \delta \rho_m= F\cdot (\delta \mu^{(P)}- \delta \mu \vert_{\delta \rho_m=0})=\delta \rho_{tot} - \delta \rho_{tot}\vert_{\delta \rho_m=0} \label{drhom} \ee where $\delta \mu^{(P)}$ is the total effective density perturbation as obtained from the Poisson Eq.~ (\ref{poisson}) and $\delta \mu \vert_{\delta \rho_m=0}$ is obtained from Eq.~ (\ref{dmu2}). The normalized matter density perturbation is obtained in terms of the field and metric perturbations as
\be \delta_m \equiv \frac{\delta \rho_m}{\rho_{tot}}=\frac{\delta \mu^{(P)}- \delta \mu \vert_{\delta \rho_m=0}}{\mu} \label{drhomrho} \ee
where $\mu$ is obtained from Eq.~ (\ref{mu}).

Alternatively, the matter density perturbation,  when normalized with respect to $\rho_m$, i.e. $\widehat{\delta}_m=\delta\rho_m/\rho_m$, may be obtained on all scales by solving the differential equation \cite{Copeland:2006wr}
\be {\ddot {\widehat{\delta}}_m}+2 H {\dot {\widehat{\delta}}_m} + \frac{k^2}{a^2}\left(\psi-\frac{F_{,\Phi}}{F}\delta \Phi\right)-3({\ddot \psi}+2H{\dot \psi})=0 \label{matperteq} \ee with appropriate initial conditions.

Even though the system (\ref{perts1}), (\ref{perts2}) for the evolution of metric and field perturbations on a scale $k$ can only be solved numerically, there are several useful qualitative conclusions that can be obtained by considering appropriate limits of the scale $k$ in these equations. There are four scales involved in Eqs.~(\ref{perts1}), (\ref{perts2}):
\begin{itemize}
\item  The physical scale $\frac{k}{a}$ of the perturbations.
\item The Hubble expansion rate $H$.
\item The mass scale of the potential $U_{,\Phi \Phi}^{1/2}$.
\item The shifted ST perturbation scale $F_{,\Phi}^{1/2}\frac{k}{a}$
\end{itemize}
In addition, in a cosmologically interesting setup, the expansion of the universe is driving the time evolution of every physical quantity $f$. Therefore, $|{\dot f}|\simeq H|f|$. For scalar fields that can play a role in the present accelerating expansion of the universe we require $U_{,\Phi \Phi}^{1/2}\simeq H$. Thus each term in Eqs.~(\ref{perts1}), (\ref{perts2}) is determined by one of the three scales: $\frac{k}{a}$, $H$, $F_{,\Phi}^{1/2}\frac{k}{a}$. By identifying the terms that dominate in each range of perturbation scales $\frac{k}{a}$ we may simplify the perturbation equations and obtain approximate solutions for the corresponding range of scales. We consider the following ranges of perturbation scales:
\begin{itemize}
\item Sub-Hubble ST scales: $\frac{k}{a}\simeq F_{,\Phi}^{1/2}\frac{k}{a} \gg H$, $F_{,\Phi}\gtrsim 1$. In this case the perturbation scale and the shifted ST perturbation scale are of the same order. Ignoring subdominant terms, Eq.~ (\ref{perts2}) becomes \be \delta \Phi \simeq (\phi-2\psi) F_{,\Phi} \label{shdphi1} \ee Using Eq.~ (\ref{psiphicon}) to express $\phi$ in terms of $\psi$ we find  \be \delta \Phi\simeq -\psi \frac{F F_{,\Phi}}{F+F_{,\Phi}^2} \label{shdphi2} \ee Thus, on sub-Hubble scales the field perturbations $\delta \Phi$ are independent of the scale $k$ and can be a significant fraction of the total energy perturbations as demonstrated in the next section. The corresponding behavior in GR is very different. Setting $F=1$ and $\psi = \phi$ in (\ref{perts2}) we obtain \be \delta \Phi \simeq A\frac{a^2H^2}{k^2}(\Phi-\Phi_i)\,\psi\rightarrow 0\quad \label{dphigr}\ee where we have used $\dot{\Phi}\simeq (\Phi-\Phi_i)\; H$ and where $A$ is a proportionality factor necessary to fit the numerical solution. As expected, the above result implies that the scalar field perturbations are negligible in GR on sub-Hubble scales.

    The field density perturbations on these scales are obtained from Eq.~ (\ref{dmu2}) by setting $\delta \rho_m=0$ and considering only the scale dependent dominant term. We thus find
    \be \delta \rho_\Phi\simeq -\frac{k^2}{a^2} F_{,\Phi}\delta \Phi = \frac{k^2}{a^2}\psi \frac{F\;F_{,\Phi}^2}{F+F_{,\Phi}^2}\label{shphipert} \ee
    The corresponding matter perturbation is obtained using Eqs.~(\ref{poisson}) and (\ref{shphipert}) as \be \delta \rho_m =\delta \rho_{tot} - \delta \rho_\Phi \simeq -\frac{k^2}{a^2}\psi F \(\frac{F_{,\Phi}^2}{F+F_{,\Phi}^2}+2\) \label{shdrm} \ee
      The sub-Hubble ratio $\frac{\delta \rho_\Phi}{\delta \rho_m}$ is therefore obtained as
      \be \frac{\delta \rho_\Phi}{\delta \rho_m}=\frac{\delta_\Phi}{\delta_m} \simeq -\frac{F_{,\Phi}^2}{3F_{,\Phi}^2 +2F}\label{shdrat} \ee
This scale independence of the ratio $\frac{\delta \rho_\Phi}{\delta \rho_m}$ on sub-Hubble scales indicates that the effective speed of sound for ST dark energy perturbations is $c_{s\Phi}=0$. The fact that $\frac{\delta \rho_\Phi}{\delta \rho_m}<0$ indicates an interesting autocorrelation between dark matter and dark energy perturbations which is also confirmed numerically in the next section.

      It is also straightforward to derive the equation for the evolution of matter density perturbations on sub-Hubble scales in ST theories. Eq.~ (\ref{matperteq}) on sub-Hubble scales takes the form
      \be {\ddot {\widehat\delta}_m}+2 H {\dot {\widehat\delta}_m} + \frac{k^2}{a^2}\left(\psi-\frac{F_{,\Phi}}{F}\delta \Phi\right)=0 \label{matperteqsh} \ee where we have $\delta \Phi$ given by Eq.~ (\ref{shdphi2}).   Eliminating $\frac{k^2}{a^2} \psi$ in Eq.~ (\ref{matperteqsh}) using Eq.~ (\ref{shdrm}) we obtain\cite{Boisseau:2000pr} \be {\ddot {\widehat{\delta}}_m}+2 H {\dot {\widehat{\delta}}_m} -\frac{\rho_m {\widehat{\delta}}_m}{2}\frac{1}{F} \frac{2F+4F_{,\Phi}^2}{2F + 3 F_{,\Phi}^2}=0 \label{matperteqsh2} \ee which provides the evolution of matter perturbations on sub-Hubble scales \cite{Boisseau:2000pr}. It may be shown that \be \frac{1}{F} \frac{2F+4F_{,\Phi}^2}{2F + 3 F_{,\Phi}^2}=\frac{G_{\rm eff}}{G} \label{geffg} \ee where $G_{\rm eff}$ is the effective gravitational constant in Cavendish-like experiments in the context of ST theories \cite{EspositoFarese:2000ij}. Therefore, Eq.~ (\ref{matperteqsh2}) has the anticipated scale independent form (as in the case of GR) but Newton's constant $G$ has been replaced by the effective ST gravitational constant $G_{\rm eff}$ (remember that we have set $8\pi G=1$).
\item Sub-Hubble GR scales:  $\frac{k}{a}\gg H\gg F_{,\Phi}^{1/2}\frac{k}{a}$, $F_{,\Phi}\ll 1$. If $F_{,\Phi}\ll 1$, there is a range of sub-Hubble  scales corresponding to \be F_{,\Phi}\ll\frac{H^2 a^2}{k^2} \ll 1 \label{grscrang} \ee for which the terms depending on the non-minimal coupling in Eq.~ (\ref{perts2}) are negligible compared to all other terms. For this range of sub-Hubble scales the scalar field perturbations are negligible, scale dependent and behave as in the case of GR (Eq.~ (\ref{dphigr})). On small enough scales however ie \be \frac{H^2 a^2}{k^2} \ll F_{,\Phi}\ll 1 \label{stscrang} \ee we reobtain the ST scale independent behavior of Eq.~ (\ref{shdphi2}) (see Fig. \ref{figratio} lower pannel).
\item Super-Hubble scales: $\frac{k}{a}\ll H$. In this case we may ignore the scale dependent terms in Eq.~ (\ref{perts2}) to obtain $\delta \Phi \sim \psi$. Clearly, there is no scale dependence for the perturbations on super-Hubble scales. Similarly, for matter perturbations we find (using Eq.~ (\ref{matperteq})) $\delta_m\sim \psi$ ie these perturbations are scale independent.
\end{itemize}
The above qualitative features of the cosmological perturbations will be confirmed and extended by the detailed numerical derivation of the perturbations that is presented in the next section.

\section{Numerical Solution}
The first step towards the numerical solution of the systems for the background evolution (Eqs.~(\ref{stbh1})-(\ref{stbmat})) and for the perturbations (Eqs.~(\ref{perts1}), (\ref{perts2})) is a proper rescaling to convert dimensional quantities to dimensionless ones. We use the following definitions of dimensionless (barred) quantities:
\ba
t&=&H_{\rm ref}^{-1} {\bar t} \\
H&=&H_{\rm ref} {\bar H} \\
k&=&H_{\rm ref} {\bar k} \\
\psi &=& \psi_i {\bar \psi} \label{srcpsi}\\
\delta \Phi &=& \psi_i {\bar{\delta \Phi}} \label{rscphi} \\
\rho_m &=& \frac{H_{\rm ref}^2}{8\pi G} {\bar \rho}_m = H_{\rm ref}^2 {\bar \rho}_m \\
U&=& \frac{H_{\rm ref}^2}{8\pi G} {\bar U} = H_{\rm ref}^2 {\bar U} \label{rescale1} \ea
where we have set $8\pi G=1$ and $H_{\rm ref}$ is a reference expansion rate used for the rescaling. Also $\psi_i$ is the initial value (at recombination) of the metric perturbation $\psi$.\footnote{The rescaling $\psi = \psi_i {\bar \psi}$ allows to set $\bar \psi=1$ in the initial conditions. Also note that $\psi_i=\psi(k,t_i)$ is in general scale dependent.} Using these rescalings, the background and perturbation equations are expressed in terms of the dimensionless barred quantities with no other changes (the form of all the equations remains the same). In what follows we omit the bar (${\bar{...}}$) but we refer to the rescaled quantities.

\begin{center}
\begin{figure}[htbp]
\epsfig{file=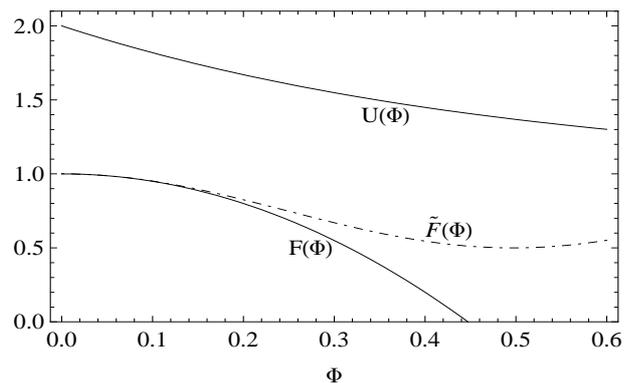,width=8cm,height=5cm}\caption{The potentials $F(\Phi)$, $\widetilde{F}(\Phi)$ and $U(\Phi)$ for the parameter values $\lambda=1$, $\lambda_f=5$, $\beta=0.5$ used in the numerical solution. In order to approximate $\widetilde{F}$ by Eq.~(\ref{fphi2}), in our numerical analysis we take the initial field value $\Phi_i\simeq 0.1$. }\label{figpot}
\end{figure}
\end{center}

The next step is to define specific forms for the potentials $F(\Phi)$ and $U(\Phi)$ to be used in the numerical solution. We consider the following form for $F$ and $U$: \ba
{\tilde F}(\Phi)&\equiv&\beta+(1-\beta )\cos^2\left(\sqrt{\widetilde{\lambda_f}}\Phi\right)\,\,,\,\,\tilde{\lambda}_f>0\,. \label{fphi}\\
U(\Phi)&=& 1 + \exp(-\lambda\Phi) \label{uphi} \ea
This form of $F(\Phi)$ is consistent with solar system tests for $\Phi\simeq 0$. Indeed solar system constraints of ST theories
imply that\cite{solsyst} \be \left.\frac{F_{,\Phi}^2}{F}\right\vert_{t=t_0} <10^{-4} \label{stconstr} \ee where $t_0$ refers to the present time. An additional advantage of this form of $F(\Phi)$, $U(\Phi)$ is that they are similar to the corresponding functions that were reconstructed \cite{Perivolaropoulos:2005yv} from the best fit expansion rate in the context of recent SnIa data while $F(\Phi)$ never becomes negative leading to instabilities. If $\Phi$ remains close to 0 during its evolution, as in the cases we consider, then we may expand around $\Phi=0$ and keep terms up to $\Phi^2$ in Eq.~(\ref{fphi}). Therefore, in our numerical analysis we use the simple generic form
\be
F(\Phi)=1-\lambda_f\Phi^2 \label{fphi2}
\ee
where $\lambda_f=(1-\beta)\widetilde{\lambda_f}<0$. The forms of $F(\Phi)$ ((\ref{fphi}) and (\ref{fphi2})) and $U(\Phi)$ (\ref{uphi}) for specific parameter values are shown are shown in Fig. \ref{figpot}. Notice that for $\Phi\lesssim 0.2$ the two forms of $F(\Phi)$ are practically identical.

We next solve the system (\ref{stbh1}),(\ref{stbphi}),(\ref{stbmat}) (after substituting ${\dot H}$ in (\ref{stbphi}) from (\ref{stbh1})) to determine the background evolution in the context of the above potentials. The initial time we consider corresponds to recombination ($a(t_i)=\frac{1}{1+z_i}= \frac{1}{1001}$) and is obtained by solving the Friedman equation ${\dot a}^2=\frac{\omm}{a(t)}$ in the matter era as \be t_i = \( \frac{4}{9\omm (1+z_i)^3} \)^{1/2}\,. \label{ti} \ee where we have set $H_{\rm ref}\simeq H_0$. The final time $t_f$ of the solution is set to  $t_f=5$ a value beyond the present $t_0$. The present time $t_0$ is determined after the solution of the system by demanding that $\omms(t_0)=\frac{\rho_m(t_0)}{3H(t_0)^2}=\omm=0.3$ ($H(t)\equiv \frac{{\dot a}(t)}{a(t)}$).

We use the following initial conditions at $t_i$ corresponding to recombination
\be \rho_m(t_i)= \frac{4}{3t_i^2} \label{rhoi} \ee obtained by assuming a matter era at early times ($H(t_i)^2 = \(\frac{2}{3t_i}\)^2=\frac{\rho_m(t_i)}{3}$), \ba & &\Phi(t_i)=0.12 \label{phiinit1} \\ & & {\dot \Phi}(t_i)=10^{-5} \label{phiinit2} \\ & & a(t_i)= \frac{1}{1+z_i} = \frac{1}{1001} \label{ainit} \ea corresponding to initial conditions at recombination close to GR. We have checked that our results are robust with respect to reasonable changes of the above initial conditions.  The parameters that need to be fixed for the solution of the background system (\ref{stbh1}),(\ref{stbphi}),(\ref{stbmat}) are $\lambda$, $\lambda_f$, $\omm$. In most solutions discussed in this section we set $\omm=0.3$ and use the potentials (\ref{uphi}), (\ref{fphi2}) with $\lambda=2$, $\lambda_f=5$ (ST cosmology) or $\lambda_f=0$ (minimally coupled quintessence).

For the background system we find two solutions and select the one that is well behaved (the other diverges and is discarded). We also rescale the scale factor by its value at the present time ($a(t)\rightarrow \frac{a(t)}{a(t_0)}$) so that $a(t_0)=1$. This rescaling shifts somewhat the initial redshift $z_i$ of our solution and the corresponding initial time given by Eq.~ (\ref{ti}). We use the function $t(z)$ obtained by solving numerically the equation $\frac{1}{a(t)}=1+z$, to find the value of the initial time at recombination corresponding to the rescaled solution $a(t)$. This function is also used to convert plots vs time to plots vs redshift. Similarly, the inversion of the equation $a=a(t)$ provides the function $t(a)$ used for the conversion of plots of quantities vs time provided by the background solution to plots vs the scale factor $a$.

\begin{center}
\begin{figure}[htbp]
\epsfig{file=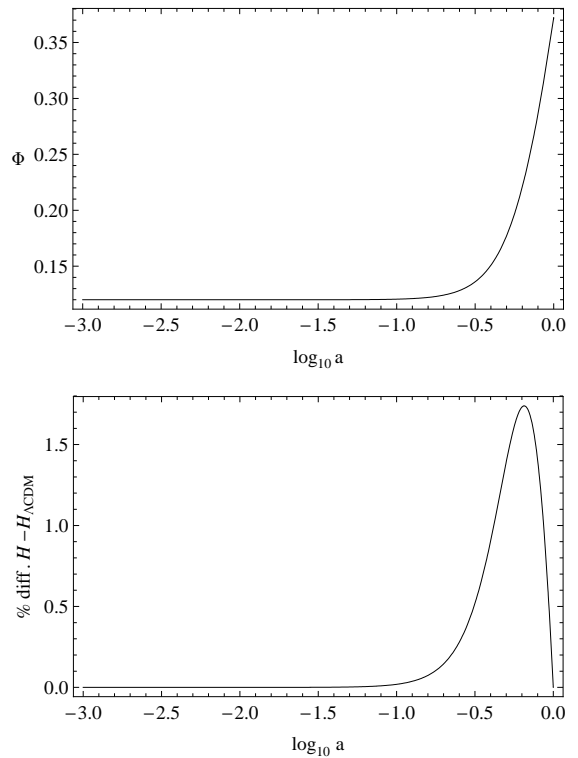,width=8.5cm}\caption{Background field dynamics for minimally coupled quintessence (top panel) and $\%$ difference between $H$ and $H_{{\Lambda}\textrm{CDM}}$ (bottom panel). Due to cosmic friction the field remains frozen at early times. Deviations from the $\Lambda$CDM expansion rate arise due to the field dynamics. During the evolution $F_{,\Phi}=0$ and $F=1$.}\label{figquint}
\end{figure}
\end{center}

We have verified that our solution reduces to \lcdm for $\lambda=\lambda_f=0$ as expected. Deviations from \lcdm are observed as we increase the value of $\lambda$, thus steepening the potential $U$ and giving dynamics to $\Phi$. Setting $\lambda_f=0$, we construct the numerical solution for the minimally coupled quintessence field and the corresponding deviation from the $\Lambda$CDM expansion rate (Fig.~\ref{figquint}). For the potential in Eq.~(\ref{uphi}) we have checked that deviations from the $\Lambda$CDM expansion rate are less than $3.5\%$ for any value of the parameter $\lambda$ with fixed value of initial condition $\Phi(t_i)\simeq 0.1$. At early times the scalar field dynamics is frozen due to cosmic friction, hence no deviation from $\Lambda$CDM arises. At late times, the constant part of the potential (\ref{uphi}) prohibits a large deviation from the \lcdm expansion rate despite of the somewhat increased kinetic energy of $\Phi$.

In Fig.~\ref{bgrl01} we plot the background dynamics for a non-minimally coupled quintessence field $\Phi$ setting $\lambda_f=5$ and $\lambda=0.1$. In ST gravity the scalar field is moving under the influence of the effective potential  (see Eq.~(\ref{stbphi}), (\ref{ricci})) \be U_{\rm eff}(\Phi)=U(\Phi)-\frac12 RF(\Phi)\,. \label{effpot}\ee Assuming that $F\sim O(1)$ at all times this implies that at early times $U_{\rm eff}(\Phi)$ is dominated by the second term. The effective mass of the field at early times is then $m_{\rm eff}\sim (-RF_{,\Phi\Phi})^{1/2}\sim\lambda_f^{1/2} H$. As a result, for $\lambda_f\gtrsim1$ and $\Phi_i\neq0$ the scalar field is dynamically rapidly driven to its value corresponding to GR ($\Phi=0$) and remains performing oscillations of decreasing amplitude around the attractor $\Phi=0$ as the Universe expands. If $\Phi_i=0$ then the field remains at $\Phi=0$ ($F\simeq 1$) until $U_{\rm eff}(\Phi)\simeq U(\Phi)$. Therefore, $\Phi$ behaves as minimally coupled quintessence at early times and as non-minimally coupled quintessence at late times when the field is driven away from $\Phi=0$. Since in the case shown in Fig. \ref{bgrl01} we chose $\lambda=0.1$, the scalar $\Phi$ is not significantly driven away from 0, hence the present value of $F_{,\Phi}$ remains small enough to satisfy the solar system constraints (Eq.~(\ref{stconstr})) since $\frac{F_{,\Phi}^2}{F}\vert_{t=t_0} \simeq 3\times10^{-4}$. Thus, a small slope of the potential $U(\Phi)$ is sufficient to secure that solar system constraints will be satisfied at present.

Deviations with respect to the \lcdm expansion rate can be significant at early times if the field begins far from $\Phi=0$. Therefore, large deviations from \lcdm at early times can be avoided simply by tuning $\Phi_i$ so that $F(\Phi_i)\simeq1$. With such initial conditions, the system approaches GR while the Hubble expansion rate becomes practically identical to \lcdm at late times. When U starts dictating the field dynamics (late time evolution), such deviations are kept minimal by assuming a small slope of the potential $U(\Phi)$ ($\lambda=0.1$, Fig. \ref{bgrl01}). As we increase the slope of the potential by increasing $\lambda$ ($\lambda=2$, Fig. \ref{figbackgr2})  we find small deviations (less than $1\%$) from the \lcdm expansion rate at late times. Even for larger values of $\lambda$ the deviations at late times are always below $3\%$ but the dynamics of the scalar field $\Phi$ is more interesting. In particular for steeper potentials $U$ ($\lambda=2$) {\it the dynamical evolution of $\Phi$ leads to an amplified value of $F_{,\Phi}^2$ at redshifts $z\lesssim O(1)$}. For example, in Fig.~\ref{figbackgr2} we plot the background dynamics with $\lambda=2$. In this case $\Phi$ has significant late time evolution. This leads to an increased value of $\frac{F_{,\Phi}^2}{F} \simeq 10^{-1}$ violating solar system constraints (Eq.~ {\ref{stconstr}) but not cosmological constraints \cite{Umezu:2005ee} \be \left.\frac{F_{,\Phi}^2}{F}\right\vert_{t=t_0} \lesssim O(1)\,. \label{cmconstr} \ee Thus, in the context of a chameleon mechanism the increased value of $F_{,\Phi}^2$ has the potential of being consistent with observational constraints. Furthermore, and as discussed in the previous section (Eq.~ (\ref{shdrat})), the value of $F_{,\Phi}^2$ determines the significance of the dark energy density perturbations compared to those of matter. We therefore anticipate \textit{amplified dark energy perturbations} (compared to GR) when the dynamics of $\Phi$ is turned on by increasing the value of $\lambda$.

\begin{widetext}
\begin{center}
\begin{figure}[htbp]
\epsfig{file=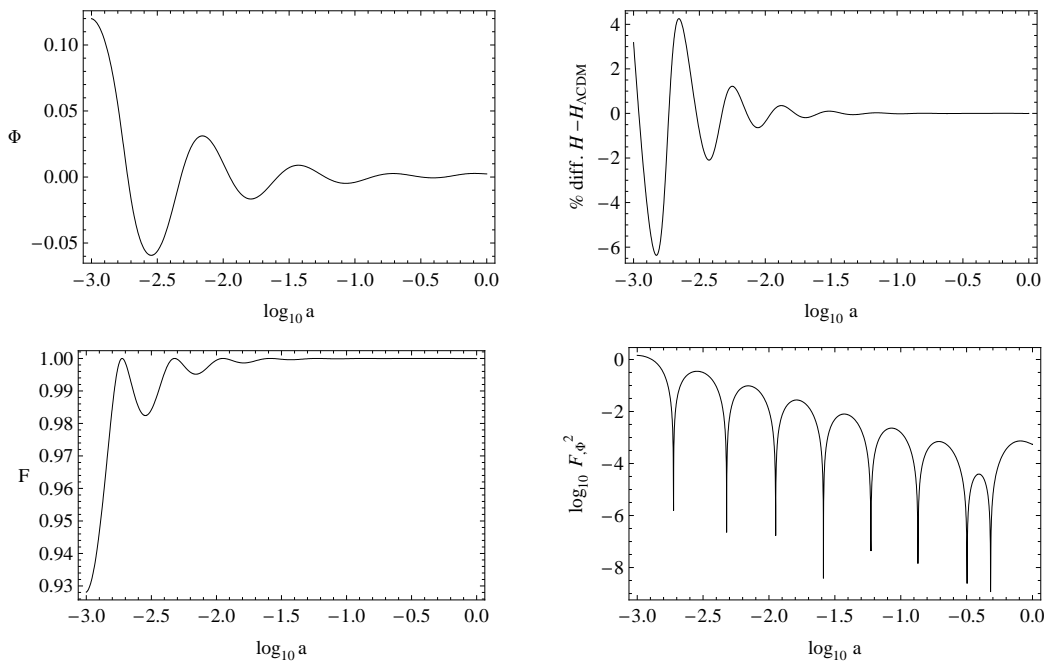,width=14cm}\caption{Background dynamics with parameters $\lambda_f=5$ and $\lambda=0.1$. From left to right and from top to bottom: 1. Field oscillations about $\Phi=0$; 2. $\%$ difference between $H$ and $H_\textrm{\lcdm}$; 3. Evolution of $F(\Phi)$;  4. $F_{,\Phi}^2$ ($\simeq 4\times10^{-4}$ at present).}\label{bgrl01}
\end{figure}
\end{center}
\end{widetext}

\begin{widetext}
\begin{center}
\begin{figure}[htbp]
\epsfig{file=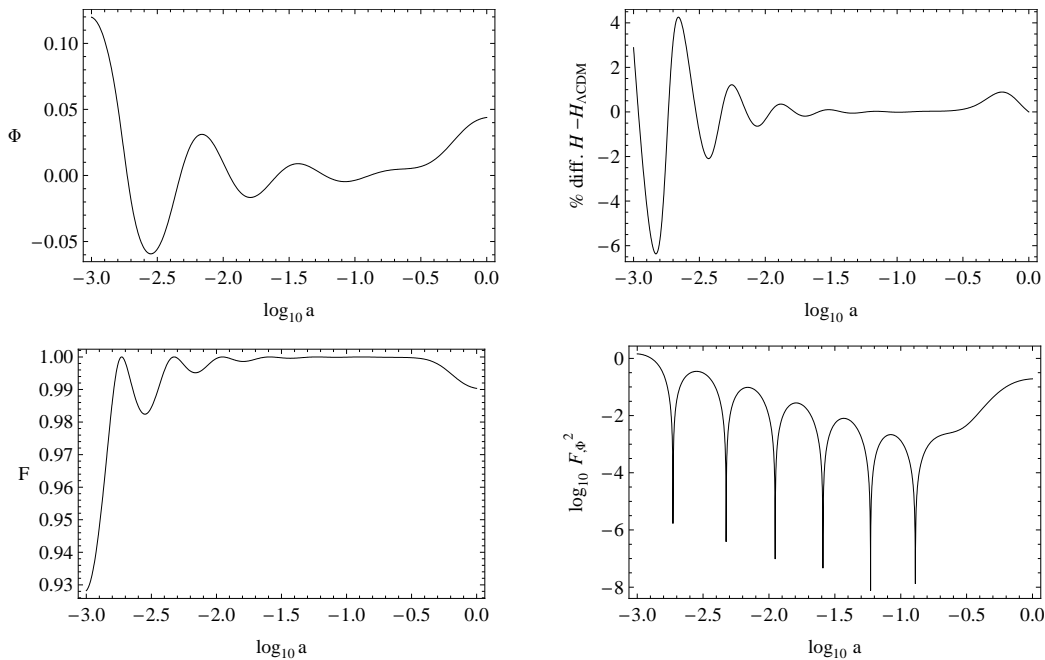,width=14cm}\caption{Background dynamics giving rise to the field contrasts depicted in Figs 6-11. From left to right and from top to bottom: 1. Field oscillations about $\Phi=0$; 2. $\%$ difference between $H$ and $H_\textrm{\lcdm}$; 3. Evolution of $F(\Phi)$;  4. $F_{,\Phi}^2$ ($\simeq10^{-1}$ at present). The parameters are $\lambda_f=5$ and $\lambda=2$.}\label{figbackgr2}
\end{figure}
\end{center}
\end{widetext}

A remarkable feature distinguishing minimally coupled quintessence from the non-minimally coupled one is that the latter is able to cross the phantom divide line \cite{Hu:2004kh,Perivolaropoulos:2005yv,Nesseris:2006er} corresponding to an effective dark energy equation of state $w_{\rm eff}=-1$. In general, the effective equation of state for the scalar field $\Phi$ is given by (see Eqs.~(\ref{stbh1}), (\ref{stbh2}))
\be
w_\Phi=\frac{\frac12\dot{\Phi}^2-U(\Phi)+\ddot{F}+2H\dot{F}}{\frac12\dot{\Phi}^2
+U(\Phi)-3H\dot{F}}\,.
\ee
In Fig.~\ref{figEOS} we show the evolution of the equation of state parameter for minimally ($\lambda_f=0$, solid line) and non-minimally ($\lambda_f=2$, dashed line) coupled quintessence, corresponding to GR and ST gravity respectively. At early times the background field oscillations give rise to divergences in the EOS parameter of $\Phi$ but such divergences do not reflect on the Hubble expansion rate. At late times, the scalar potential $U(\Phi)$ becomes relevant for the field dynamics, the field starts growing and the EOS parameter oscillates around the phantom divide line $w=-1$. Crossing of the phantom divide line is allowed by all current cosmological observations and is in fact favored by some of them \cite{Nesseris:2006er}. This behavior is characteristic of ST gravities and cannot be achieved in minimally coupled quintessence \cite{Vikman:2004dc}.

\begin{center}
\begin{figure}[htbp]
\epsfig{file=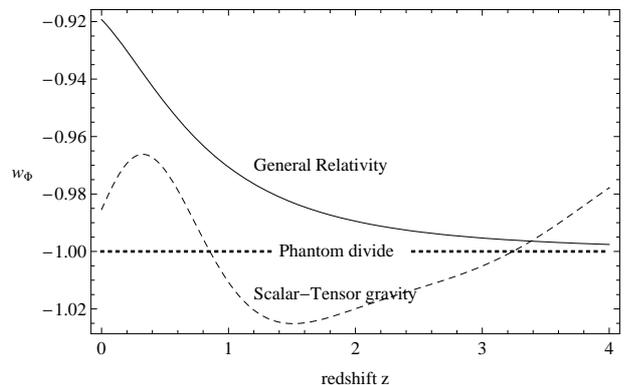,width=8.0cm}\caption{Equation of state parameter (EOS) w$_\Phi$ for the background field $\Phi$ as obtained in General Relativity ($\lambda_f=0$, $\lambda=2$, solid line) and Scalar-Tensor gravity ($\lambda_f=5$, $\lambda=2$, dashed line). In Scalar-Tensor gravity the non-minimal coupling of the field results in the crossing of the phantom divide line (in red); for minimally coupled quintessence such crossing does not occur.}\label{figEOS}
\end{figure}
\end{center}

We now use the background solution in Eqs.~(\ref{perts1}), (\ref{perts2}) to find the evolution of perturbations. We use initial conditions corresponding to matter era in GR. Taking also into account the rescalings (\ref{srcpsi}), (\ref{rscphi}), the initial conditions used for the solution of (\ref{perts1}), (\ref{perts2}) are of the form\footnote{Note that $\psi(k,t_i)$ is actually $\bar\psi(k,t_i)\equiv \frac{\psi(k,t_i)}{\psi(k,t_i)}=1$ (rescaled by the initial gravitational potential). The initial gravitational potential corresponding to a scale invariant Harrison-Zeldovich spectrum obeys $\phi(k,t_i)^2 k^4\sim \delta_m(k,t_i)^2\sim k$ and therefore $\phi(k,t_i)^2\sim k^{-3}$.}
\ba \delta \Phi (k,t_i)= {\dot {\delta \Phi}}(k,t_i)&=&0 \label{dfi} \\
\psi(k,t_i)=1 \; ; \; {\dot \psi}(k,t_i) &=& 0 \label{dpsii} \ea
In order to find the value of the rescaled wavenumber ${\bar k}$ used in the perturbation Eqs.~(\ref{perts1}), (\ref{perts2}) corresponding to a given physical scale $\lambda_p \; \textrm{h$^{-1}$Mpc}$ we express ${\bar k}$ as follows:
\be {\bar k}= \frac{2\pi}{\lambda_p \; \textrm{h$^{-1}$Mpc}}\frac{c}{H_{\rm ref}}=\frac{2\pi}{\lambda_p \; \textrm{h$^{-1}$Mpc}}\frac{3\times 10^5 \textrm{km\,sec$^{-1}$}}{100 \textrm{h km\,sec$^{-1}$}}{\bar H}_0 \label{barkder} \ee where $c$ is the velocity of light and ${\bar H}_0=\frac{H_0}{H_{\rm ref}}$ is the present value of the Hubble parameter as provided by the solution of the rescaled background system (\ref{stbh1}),(\ref{stbphi}),(\ref{stbmat}).

The evolution of the field $\delta \Phi$ and metric perturbations $\psi$ vs the scale factor $a$ (in logarithmic scale) for a perturbation of wavelength $\lambda_p=30$ h$^{-1}$Mpc are shown in Figs.~\ref{figdeltaphi} and \ref{figpsi}. In the top panel of Fig. \ref{figdeltaphi} we plot the numerical solution for the evolution of $\delta\Phi$ as obtained in GR (solid line) and its predicted value on subhorizon scales (dashed line), given by Eq.~(\ref{dphigr}).
We set $A\simeq25$ in Eq.~(\ref{dphigr}) in order to match the numerical solution. The bottom panel shows the numerical solution for the evolution of $\delta\Phi$ in ST gravity (solid line) along with its predicted value on subhorizon ST scales (dashed line), given by Eq.~(\ref{shdphi2}). As predicted by the analytical expression (\ref{shdphi2}), the field perturbation $\delta \Phi$ is a product of two oscillating modes: $\psi$ whose oscillations on sub-Hubble scales are driven by the term $\frac{k^2}{a^2} F_{,\Phi} \delta \Phi$ (see Eqs.~(\ref{perts1}), (\ref{dpdef})) with frequency \be \omega_\psi = \frac{k}{a H} m_{\Phi {\rm eff}} \simeq \frac{k}{a H}  \lambda_f^{1/2} H \label{ompsi} \ee and $F_{,\Phi}$ with frequency \be \omega_\Phi \simeq m_{\Phi {\rm eff}} \simeq \lambda_f^{1/2} H \label{omphi} \ee This superposition of high and low frequency modes ($\omega_\psi \gg \omega_\phi$ on sub-Hubble scales) manifests itself in Fig. \ref{figdeltaphi}.

\begin{center}
\begin{figure}[htbp]
\epsfig{file=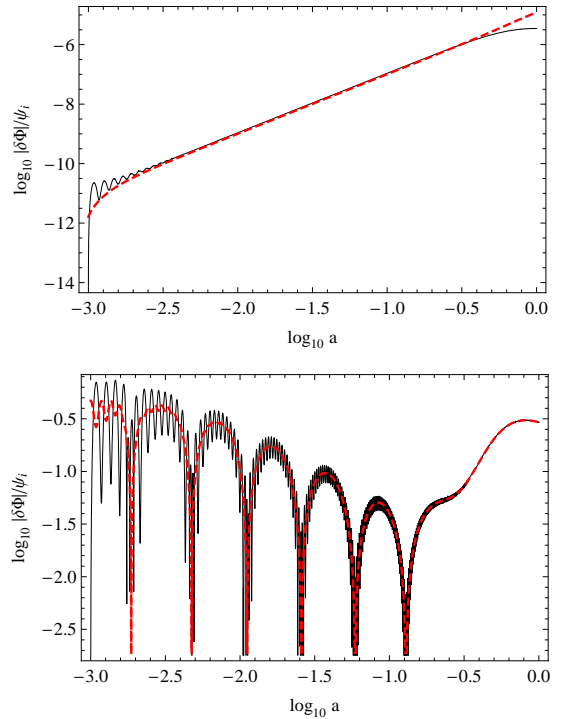,width=8.0cm}\caption{Evolution of the field perturbation $\delta\Phi$ as obtained in General Relativity (top panel) and Scalar-Tensor gravity (bottom panel) for the scale $\lambda_p=30$ h$^{-1}$Mpc. The numerical solution is the solid line and the analytical approximation on subhorizon ST and GR scales is the dashed lines. Oscillations in the background field $\Phi$ induce oscillations in $\delta\Phi$ through $F_{,\Phi}\propto\Phi$. We use $\lambda_f=5$ for ST gravity and $\lambda=2$ in both cases. The spikes correspond to changes of sign of $\delta \Phi$.}\label{figdeltaphi}
\end{figure}
\end{center}

\begin{center}
\begin{figure}[htbp]
\epsfig{file=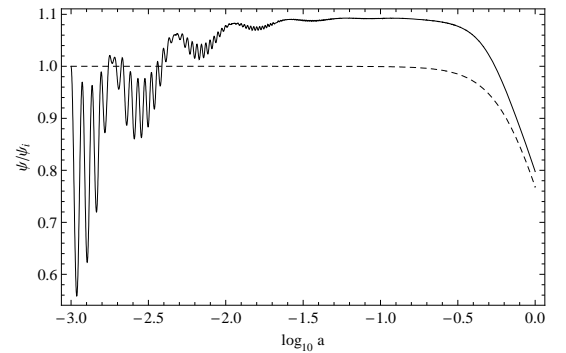,width=7cm}\caption{Evolution of the metric perturbation $\psi$ as obtained in General Relativity ($\lambda_f=0$, dashed line) and ST gravity ($\lambda_f=5$, solid line) for the scale $\lambda_p=30$ h$^{-1}$Mpc. We use $\lambda=2$ in both cases.}\label{figpsi}
\end{figure}
\end{center}

\begin{center}
\begin{figure}[htbp]
\epsfig{file=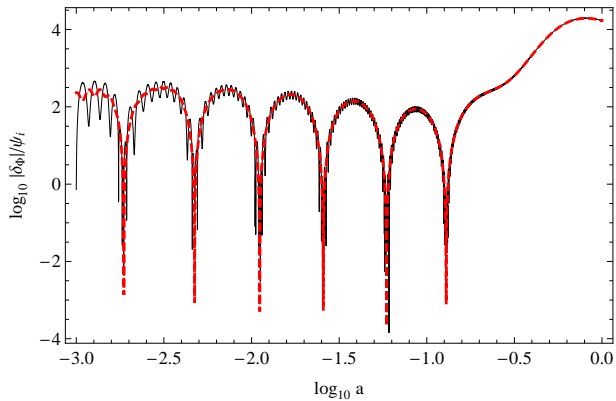,width=8.0cm}\caption{Evolution of the density contrast $\delta_\Phi$ in ST gravity for a perturbation of wavelength \mbox{$\lambda_p=30$ h$^{-1}$Mpc}. The numerical solution is the solid line and the theoretical prediction, obtained using Eqs.~(\ref{shdphi2}) and (\ref{shphipert}), is the dashed line. The sign of $\delta_\Phi$ is negative, hence an underdensity in the DE fluid is created. We use the values $\lambda_f=5$ and $\lambda=2$.}\label{figdeltaphi2}
\end{figure}
\end{center}

\begin{center}
\begin{figure}[htbp]
\epsfig{file=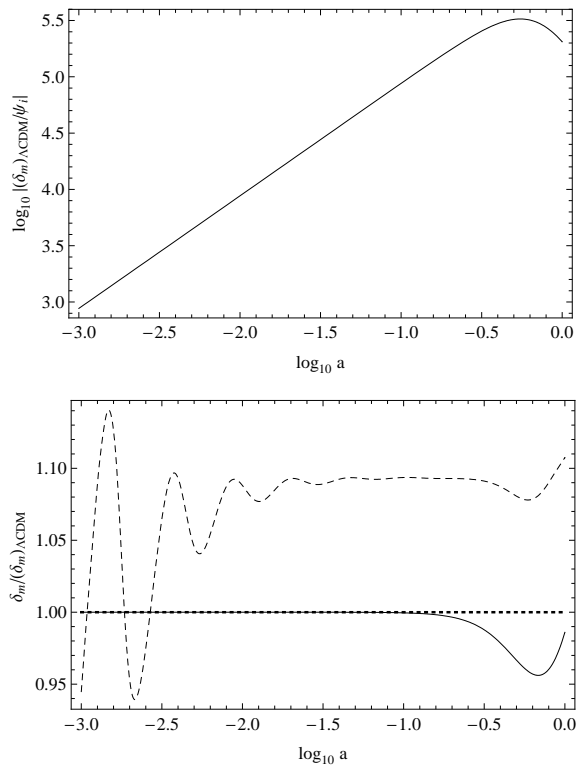,width=8.5cm}\caption{Top panel: Matter density perturbation ${\delta_m}_{\Lambda CDM}$ in \lcdm. Bottom panel: The
ratio $\frac{\delta_m}{{\delta_m}_{\Lambda CDM}}$ where $\delta_m$ corresponds to ST gravity (dashed line),
minimally coupled quintessence in GR (solid line) and \lcdm (dotted line).}\label{figdeltam}
\end{figure}
\end{center}

In Fig.~\ref{figpsi} we plot the evolution of the metric perturbation $\psi$ with the scale factor as obtained in GR (dashed line) and in ST gravity (solid line). Since $\psi$ is driven by terms having both frequencies $\omega_\psi$ and $\omega_\phi$ (see Eqs.~(\ref{perts1}), (\ref{dpdef})) we also see in the evolution of $\psi$ a superposition of high and low frequency modes. Notice that no such oscillations are present in the GR case (dashed line in Fig. \ref{figpsi}).

In Fig.~\ref{figdeltaphi2} we plot the evolution of the density contrast $\delta_\Phi$ obtained in ST gravity with the scale factor. As expected, the field contrast shows the same oscillations as $\delta_\Phi$. Also expected is the good agreement between the numerical solution (solid line) and the analytical prediction (dashed line). The latter is obtained by using Eqs.~(\ref{phiperts}), (\ref{shphipert}) and the background solution for $\rho_{\rm tot}$ (equations \ref{mu}, \ref{rhotot}). On subhorizon ST scales we obtain a negative value of $\delta_\Phi$, as opposed to the positive value of $\delta_m$. This implies that the field and matter perturbations are anti-correlated, i.e. an overdensity in the matter gives rise to an underdensity in the scalar field $\Phi$. This result has also been obtained in the context of minimally coupled quintessence \cite{Dutta:2006pn}. In that case however it is negligible on sub-Hubble scales.

\begin{center}
\begin{figure}[htbp]
\epsfig{file=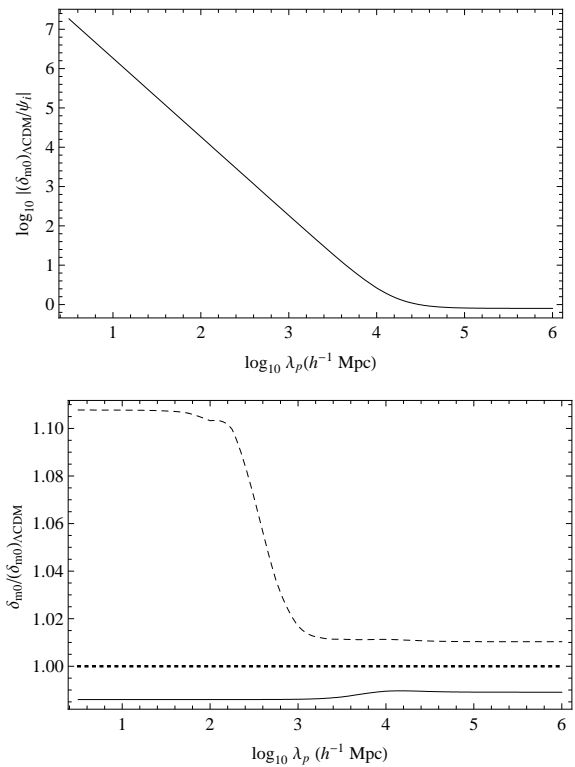,width=8.5cm}\caption{Top panel:
The  scale dependence of the rescaled present matter overdensity $\delta_{m\Lambda CDM}(k,t_0)/\psi_i$ as predicted by \lcdm. Note that $\psi_i\equiv \psi(k,t_i)$ is in general scale dependent. Bottom panel: The ratio $\frac{\delta_m(k,t_0)}{\delta_{m\Lambda CDM}(k,t_0)}$ for ST gravity (dashed line) and GR quintessence (solid line).}\label{figdeltam2}
\end{figure}
\end{center}

The evolution of matter overdensities is also affected by the introduction of a non-minimal coupling. This is demonstrated in Fig. \ref{figdeltam} where we show the evolution of $\delta_{m\Lambda CDM}$ in \lcdm on a scale of $30 {\rm h^{-1} Mpc}$ (top panel) and the ratio $\frac{\delta_m}{\delta_{m\Lambda CDM}}$ where $\delta_m$ corresponds to ST gravity (dashed line), minimally coupled quintessence in GR (solid line) and \lcdm (dotted line). Clearly, the matter overdensity is amplified by about $10\%$ in ST gravity while it is practically identical to \lcdm in GR quintessence. This amplification of $\delta_m$ in ST gravity could be made even larger at the expense of introducing more significant variation of the background $H(z)$ from \lcdm than shown in Fig. \ref{bgrl01}. In Fig. \ref{figdeltam2} we show the corresponding scale dependence of the present matter overdensity $\delta_{m\Lambda CDM}(k,t_0)$ as predicted by \lcdm (top panel) and the ratio $\frac{\delta_m(k,t_0)}{\delta_{m\Lambda CDM}(k,t_0)}$ for ST gravity (dashed line) and GR quintessence (solid line). Clearly, the $10\%$ amplification of matter perturbations in ST gravity is applicable on sub-Hubble scales while on larger scales the amplification is negligible. This mild amplification of matter perturbations may be attributed to the corresponding amplification of dark energy perturbations in these theories which also affects matter perturbations despite the predicted anti-correlation. Indeed, the dark energy void in a cluster of galaxies reduces the negative pressure inside the cluster and amplifies the gravitational collapse.

\begin{center}
\begin{figure}[htbp]
\epsfig{file=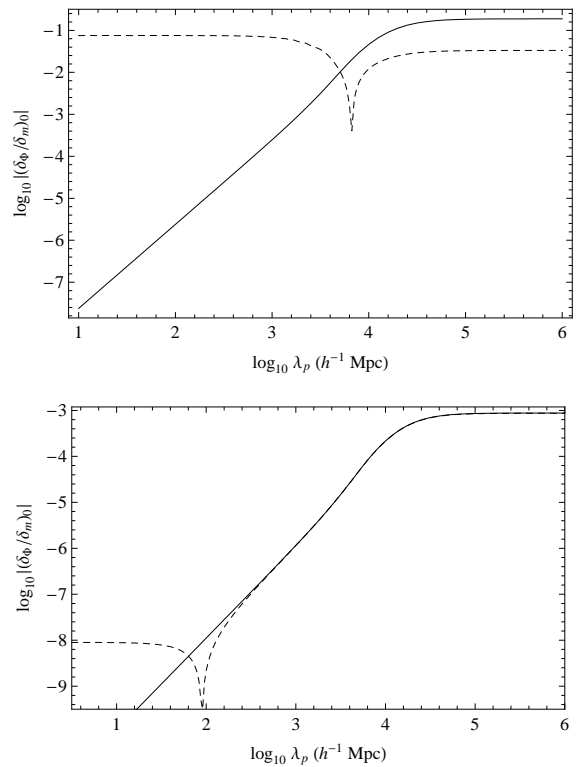,width=8.5cm}\caption{Scale dependence of the ratio $\delta_\Phi/\delta_m$ at present as obtained in GR (solid line) and ST gravity (dashed line). For the top panel we use $\lambda_f=5$ and $\lambda=2$, in which case there is no range of scales on which ST gravity behaves as GR. In  order to display such an interval of scales more clearly, in the bottom panel we set $\lambda_f=5\times10^{-4}$ and $\lambda=0.1$.}\label{figratio}
\end{figure}
\end{center}

In Fig.~\ref{figratio} we plot the scale dependence of the ratio $\delta_\Phi/\delta_m$ at present in GR (solid line) and in ST gravity (dashed line). Setting $\lambda_f=5$, $\lambda=2$, in the top panel we find that the GR solution ($\lambda_f=0$, solid line) leads to negligible dark energy perturbations on sub-Hubble scales ($\delta_\Phi \sim k^{-2}$, Eq.~ (\ref{dphigr})). In contrast, ST gravity ($\lambda_f=2$, dashed line) produces amplified, anticorrelated with matter, dark energy perturbations on sub-Hubble scales. The ratio $\frac{\delta_\Phi}{\delta_m}$ is scale independent on these scales as predicted by Eq.~ (\ref{shdrat}).  In such models where the dark energy perturbations can grow on all scales it may be shown \cite{dePutter:2010vy} that the speed of sound $c_s$ vanishes. The anti-correlation is evident by the spike of the dashed line which corresponds to a change of sign of $\delta_\Phi$ on sub-Hubble scales. For $F_{,\Phi}\ll 1$ (Fig. \ref{bgrl01}) and flat potential $U$ ($\lambda =0.1$), there is a range of intermediate sub-Hubble scales where ST perturbations behave as in the case of GR ($\delta_\Phi \sim k^{-2}$, lower panel of Fig. \ref{figratio}). This case was discussed in section 3 using analytical arguments (sub-Hubble GR scales). On super-Hubble scales we obtain the anticipated scale independence. We have checked that the agreement between the numerical result for the ratio $(\delta_\Phi/\delta_m)_0$ and the analytical predictions on sub-Hubble ST (Eq.~ (\ref{shdrat})) and GR scales is very good. This is expected in view of the good agreement between the numerical solution and analytical expressions $\delta\Phi$ both in ST gravity, Eq.~(\ref{shdphi2}), and in GR, Eq.~(\ref{dphigr}).

\section{Conclusion}
We have investigated in detail, analytically and numerically, the evolution of dark energy and matter linear density perturbations in Scalar-Tensor (ST) cosmologies. We have found that the evolution of dark energy perturbations in ST cosmologies is significantly different from the corresponding evolution in minimally coupled (GR) quintessence. In particular, our results may be summarized as follows:
\begin{itemize}
\item
For natural ($O(1)$) values of the ST Lagrangian parameters which lead to a background expansion  similar to \lcdm, ST dark energy density perturbations are amplified by a factor of about $10^6$ compared to minimally coupled quintessence perturbations on scales less than about $100{\rm h^{-1}Mpc}$ (Fig. \ref{figratio}).
\item
On sub-Hubble scales dark energy perturbations constitute a fixed fraction of about $10\%$ compared to matter density perturbations (Fig. \ref{figratio}). The fixed scale independent fraction implies that the effective speed of sound for ST dark energy is $c_{s\Phi}=0$. The corresponding fraction for minimally coupled quintessence perturbations scales as $k^{-2}$ and is about $\lesssim 10^{-4}\%$ (Fig. \ref{figratio}) corresponding to $c_{s\Phi}=1$.
\item
Scalar-Tensor dark energy density perturbations are anti-correlated with matter linear perturbations on sub-Hubble scales (Eq.~ (\ref{shdrat}) and Fig. \ref{figratio}  where $\delta_m$ and $\delta_\Phi$ are shown to have opposite signs). Thus clusters of galaxies overlap with voids of dark energy.
\item
The evolution of scalar field perturbations on sub-Hubble scales, is scale independent and involves large oscillations (Fig. \ref{figdeltaphi}) induced by the amplified effective mass of the field (Eq.~ (\ref{effpot})). This mass amplification is due to the non-minimal coupling of the field to curvature and (therefore) to matter (Eqs.~(\ref{stbh2}), (\ref{stbphi})). No such oscillations are present in minimally coupled quintessence perturbations which are suppressed on sub-Hubble scales and vary as $k^{-2}$ (Eq.~ (\ref{dphigr})).

\item
The evolution of matter density perturbations is affected by the introduction of non-minimal coupling (Figs. 9-10) and is amplified by about $10\%$ in ST cosmology compared to minimally coupled quintessence and \lcdm on sub-Hubble scales.
\item
For small values of non-minimal coupling $F_{,\Phi}$ there is a range of sub-Hubble scales where the scalar field perturbations have a scale dependence similar to the case of GR ($\sim k^{-2}$). However, even in this case, for small enough scales $\frac{k}{a}\gtrsim \frac{H}{F_{,\Phi}}$ the field perturbations become scale independent and enter the ST regime (Fig. \ref{figratio} lower panel).
\end{itemize}

These results have interesting observational consequences. In particular
\begin{itemize}
\item
{\bf Dark Matter Halo Profiles:}  \lcdm predicts shallow low concentration density dark matter halo profiles for clusters and galaxies in contrast to observations which indicate denser high concentration cluster haloes\cite{Broadhurst:2004bi}. The amplified anti-correlated with matter dark energy perturbation profiles can lead to a modification of the predicted by \lcdm dark matter halo profiles. In particular, the dark energy voids in clusters of galaxies can amplify locally dark matter clustering due to the local reduction of negative pressure in the region of the cluster. A detailed investigation of this effect would require the solution of the full coupled nonlinear system for the evolution of dark energy and dark matter perturbations under the assumption of spherical symmetry. This is a straightforward generalization of the present study.
\item
{\bf Large Scale Structure Power Spectrum $P_m(k)$:} For a non-minimal coupling $F_{,\Phi}= O(1)$ the ratio $\frac{\delta_{\Phi}^2}{\delta_m^2}\sim \frac{P_\Phi (k)}{P_m(k)}$ is scale independent for practically all sub-Hubble scales (see Fig. \ref{figratio} upper pannel). Thus it would be hard to identify a scale dependent signature of dark energy perturbations on the matter power spectrum for such values of $F_{,\Phi}$. For smaller values of the non-minimal coupling however, there will be a GR regime for large sub-Hubble scales where the dark energy perturbations are predicted to be scale dependent (Fig. \ref{figratio} lower panel) while on smaller scales we enter the ST regime where the ratio $\frac{P_\Phi (k)}{P_m(k)}$ becomes again scale independent. This transition from the GR regime on large sub-Hubble scales to the ST regime in small sub-Hubble scales may leave a trace (small glitch) on the matter power spectrum on a scale $\frac{k}{a}\simeq \frac{H}{F_{,\Phi}^{1/2}}$.
\item
{\bf Lensing by Galaxy Clusters:} The lensing properties of galaxy clusters may well be altered due to the presence of dark energy voids. It would be interesting to investigate the lensing signatures predicted by the superposition of dark energy voids on galaxy clusters.
\end{itemize}
In conclusion, the amplified and anti-correlated with matter, dark energy ST perturbations investigated in the present study provide a new direction of observational signatures for this class of modified gravity models.

{\bf Numerical Analysis Files:} The mathematica files used for the numerical analysis of this study and the production of the figures may be found at http://leandros.physics.uoi.gr/deperts/deperts.htm .

\section*{Acknowledgements}
This work was supported by the European Research and
Training Network MRTPN-CT-2006 035863-1 (UniverseNet).

\end{document}